\documentclass[11pt,onecolumn]{article}
\usepackage[top=1in, bottom=1in, left=1.25in, right=1.25in]{geometry}
\usepackage{appendix}
\usepackage{amsfonts}
\usepackage{amsmath}
\usepackage{amssymb}
\usepackage{graphicx}
\usepackage{color, soul}
\usepackage{stfloats}

\newtheorem{theorem}{Theorem}
\newtheorem{lemma}{Lemma}
\newtheorem{remark}{Remark}
\newtheorem{assumption}{Assumption}

\begin{document}
\title{Edge Balance Ratio: Power Law from Vertices to Edges in Directed Complex Network}
\author{Xiaohan Wang, Zhaoqun Chen, Pengfei Liu, and Yuantao Gu\thanks{The authors are with Department of Electronic Engineering, Tsinghua University, Beijing 100084, China. The corresponding author of this paper is Yuantao Gu (Email: gyt@tsinghua.edu.cn).}}

\date{Received Aug. 15, 2012; Accepted Jan. 31, 2013.\\\vspace{1em}
This article will appear in \textsl{IEEE Journal of Selected Topics in Signal Processing}.}

\maketitle
\begin{abstract}
Power law distribution is common in real-world networks including online social networks. Many studies on complex networks focus on the characteristics of vertices, which are always proved to follow the power law. However, few researches have been done on edges in directed networks. In this paper, edge balance ratio is firstly proposed to measure the balance property of edges in directed networks. Based on edge balance ratio, balance profile and positivity are put forward to describe the balance level of the whole network. Then the distribution of edge balance ratio is theoretically analyzed. In a directed network whose vertex in-degree follows the power law with scaling exponent $\gamma$, it is proved that the edge balance ratio follows a piecewise power law, with the scaling exponent of each section linearly dependents on $\gamma$. The theoretical analysis is verified by numerical simulations. Moreover, the theoretical analysis is confirmed by statistics of real-world online social networks, including Twitter network with 35 million users and Sina Weibo network with 110 million users.

\end{abstract}
\section{Introduction}
\subsection{Complex Network and Power Law}

Large numbers of real-world systems can be described as complex networks \cite{thestructure,complex,exploring,thestructureand}, which are represented as undirected or directed graphs. Individuals in the system are represented as vertices and interactions between them are represented as edges. The Internet \cite{onpowerlaw}, social networks \cite{social}, scientists cooperation networks \cite{howpopular}, protein interaction networks \cite{thelargescale} are several examples. In general, complex networks contain large amounts of vertices. Interactions between the vertices are neither purely regular nor purely random. Despite the different appearances of networks, the nature of many networks has similarities. Researches on complex networks try to understand the structure and behavior of networks, which are important in various areas.

Several characteristics are shared in most real-world networks. The most well-known properties are small-world \cite{collective,small} and scale-free \cite{emergence,scale,scalefree}. In a small-world network, the average distance between vertices is small and the network has a high clustering coefficient, which represents the density of triangles in the network. In a scale-free network, the degree distribution follows a power law. In other words, the probability of vertices with degree $k$ satisfies,
$$
P(k)\sim k^{-\gamma},
$$
where $\gamma$ is the scaling exponent. Most real-world networks follow the scale-free property, with the value of $\gamma$ typically satisfying $2<\gamma<3$ \cite{statistical}. It is indicated that in scale-free networks, large amounts of vertices have small degrees, while small amounts of vertices have degrees significantly larger than others. Power law distribution exists widely in nature, and attracts greatly attentions from researchers. Besides, some real-world networks have the characteristic of self-similarity \cite{selfsimilarity}, which means that part of the network is similar with the whole network. Some networks are structural \cite{topological} or hierarchical \cite{hierarchical}, and can be clustered or divided.

\subsection{Vertices and Edges}

Vertices are important components in complex networks, as the representations of individuals. Abundant researches have been done on vertices in complex networks. The degree distribution of vertices is an important property of a network. The steady-state transition probability of vertices is utilized in PageRank to measure the importance of vertices \cite{pagerank}. Traditional community detection methods are designed based on partition of vertices \cite{communitydetection}. Random walk is widely used in collecting information or recovering structures in large networks \cite{maps}.
Many efficient algorithms are proposed based on random walk to explore the network, including adaptive methods \cite{exploringcomplex}.
However, most of random walk methods are based on the properties of vertices. Vertices always attract more attentions than edges in the research on complex networks.

Edges are also important components in networks, representing the interactions between vertices. In computer and communication networks, edges represent the connections. In social networks, edges represent the relationships between users. Recently, researches on edges attract more and more attentions \cite{linkmining}. Link prediction \cite{thelink,linkprediction} in complex networks is extensively studied.
Random walk based on edges is introduced in \cite{linegraphs}, where walkers move between adjacent edges.
In \cite{linkcommunities}, hierarchical clustering on edges is used in order to detect overlapping communities in networks, which is a promising strategy to analyze graphs.
However, the researches on edges are not as many as those on vertices in complex networks. Research works on more complex cases such as directed edges in networks are still very little.

Line graph is introduced in \cite{schaum}, which transforms the edges and vertices of the network. Line graph is the graph in which the vertices represent the edges of the original graph and vertices are connected if the corresponding edges are adjacent.

Although edges of complex networks have attracted attentions from researchers and lots of work has been done, there are still a lot to be revealed, especially for directed networks. In this work, edge balance ratio is proposed as a measure of the balance property of directed edges. The distribution of edge balance ratio is studied for power law networks.

\subsection{Assortativity and Hierarchy}

Assortativity or assortative mixing is firstly proposed in \cite{assortativemixing} to quantify the mixing property in networks. Vertices in assortative networks tend to connect to the vertices with similar property, which is typically the degree of the vertex. On the contrary, in disassortative networks, high degree vertices tend to connect to low degree vertices.
Assortativity is generally studied for undirected networks, however, approaches on directed networks are introduced in \cite{mixingpatterns} and \cite{edgedirection}. In \cite{edgedirection}, four directed assortativity measures are used to quantify the correlations of combinations of in-degree and out-degree separately.
For example, the in-out assortativity represents the correlation of the in-degree of the source vertex and the out-degree of the target vertex.
Local assortativity is proposed in \cite{localassortativity} to measure the assortative level of each individual vertex in the context of the overall network.
Link assortativity is defined in \cite{linkassortativity} for directed networks to analysis the assortative property of directed edges.

Hierarchy is a critical nature of networks, especially directed networks.
A technique for inferring hierarchical structure from network data is presented in \cite{hierarchicalstructure} to explain the topological properties of networks.
A maximum likelihood based method is proposed in \cite{inferringthemaximum} to infer social hierarchy from social networks.
Various definitions of hierarchy measures are proposed for directed networks to reveal the hierarchical property \cite{hierarchymeasures, hierarchymeasure, findinghierarchy}.
In \cite{findinghierarchy}, a rank is assigned to each vertex and the basic idea is to minimize the total ``agony'' which is led by edges pointing to lower ranked vertices from higher ranked ones.
A measure of hierarchy is defined based on the minimal agony and reveals the hierarchical level of the network.

\subsection{Real-world Networks: Microblogging Networks}
Social network is a typical kind of complex network and mostly has the properties of small-world and scale-free. Microblogging network is an important kind of social network, with Twitter \cite{twitter} and Sina Weibo \cite{weibo} as typical representatives.
Analysis on Twitter and Sina Weibo has been done in \cite{whywe, whatis, followwhom}.

Twitter is a widely used online microblogging service. People can follow others they interest in on Twitter to build their personal online social relations. Twitter users can publish text messages called \emph{tweets} on their home pages and read tweets from whom they follow. One tweet is limited up to 140 characters and it can be published with links and pictures.

The social network built on Twitter is different from the one in real society because the relationships on Twitter can be unidirectional: you never need to get the person's approval when you decide to follow him. Besides, tweets always flow from the publisher to the users who follow the publisher and these users are called followers or fans. On Twitter, if one user has lots of followers, what he posts in tweets will be delivered to a wide range of users.

In China, there are many online social networking services similar to Twitter. Sina Weibo is the most famous Chinese microblogging service. \emph{Weibo} is the word of microblogging in Chinese. Sina Weibo has similar features with Twitter such as directed relationship and publishing text-message, called weibo, with limited characters. It is reported that Sina Weibo has 324 million registered users by the end of the first quarter of 2012.

Besides, there are other microblogging services worldwide such as Tumblr and Plurk. Some microblogging services are locally used with special languages supported. Qaiku was launched in Finland to be a Finnish service, while ImaHima is popular in Japan. In China, FanFou is the first Chinese microblogging service website and now more appear like Tencent Weibo and Sohu Weibo.

A microblogging network is a directed complex network. The users are represented as vertices and the relationships are represented as directed edges. If user A follows user B, there is an edge from vertex A to B in the directed graph. Each edge represents a following  relationship in microblogging networks.

\subsection{Our Work}
One main contribution of this work is to propose balance measures for directed edges and the whole network, and to discover the power law property of the edge balance ratio in a power law network. Edge balance ratio is proposed as a measure of balance level of directed edges. Balance profile is defined to describe the global balance property of the whole network. The positivity of a network is defined to reveal the positive level of it. It is theoretically analyzed that for a network with power law in-degree distribution, the edge balance ratio follows a piecewise power law distribution, and the scaling exponents are determined by the scaling exponent of in-degree. The distribution curve of edge balance ratio is wizard-hat shaped. Edge balance ratio is the property of edges, while in-degree is the property of vertices. Our work establishes the link between them. Statistics of numerical simulations and real-world datasets confirm the theoretical analysis.

The paper is organized as follows. In section II, some basic definitions including edge balance ratio and balance profile are proposed and the main contribution on the edge balance ratio is given, with theoretical analysis.  In section III, conditions and definitions are proposed to establish the network model and simulations are performed to verify the theoretical results. Statistics results on real-world network datasets are in section IV. Some discussions are in section V and the paper is concluded in section VI.

\section{Main Result on Edge Balance Ratio}

In this section, edge balance ratio is proposed as a measure of the balance level of an edge. Based on edge balance ratio, balance profile and positivity are defined as a description of the global balance property of the directed network.
The distribution of edge balance ratio for power law network is theoretically analyzed, which is the main result of this work.

\subsection{Edge Balance Ratio}
This paper focuses on the basic properties of edges. The in-degree of a vertex is the amount of edges pointing to it. If there is an edge from vertex A to B, we call vertex A the out-vertex of the edge. Correspondingly, B is called the in-vertex.
The in-vertex and out-vertex are not equivalent in a directed edge, which means edges are unbalanced in a directed network.
In order to describe the balance level of edges, \emph{edge balance ratio} is defined as one of the properties of directed edges. For a directed edge from vertex A to B, edge balance ratio $R$ is defined as
$$
R=\left\{
    \begin{array}{ll}
    \displaystyle{\frac{d_i({\rm B})}{d_i({\rm A})}}, &d_i({\rm A})\neq0; \\
    \infty, &d_i({\rm A})=0,
    \end{array} \right.
$$
where $d_i({\rm B})$ and $d_i({\rm A})$ are the in-degrees of vertices B and A.
The \emph{logarithmic edge balance ratio} is defined as the logarithm of edge balance ratio, which is more convenient to describe the characteristics of the whole network.

The edge balance ratio reflects the balance property of edges. In a common network, the in-degree of a vertex roughly
reflects its importance. In most cases, a vertex with a larger in-degree is a more important vertex in the network. An edge with a large edge balance ratio implies that the in-vertex has a much larger in-degree than the out-vertex, which implies that the in-vertex is very likely to be more important than the out-vertex. A balanced edge is one whose in-vertex and out-vertex have similar importance.

Unbalanced edges are common in real-world networks. Specially, in a directed network which follows power law, there are many edges pointing to the vertices with large in-degrees. Most of edges like these are unbalanced ones.

In microblogging networks there are abundant unbalanced edges. For instance, famous stars attract large amounts of followers. As a result, most of these edges are extremely unbalanced and edges like these accounts for a large proportion in microblogging networks.
Edge balance ratio implies the type of edges in a microblogging network. The edges with balance ratios far larger than one reflect the common following relationships, in which most uses are likely to follow users more famous than themselves. Edges with balance ratios close to one represent the relationships between friends or people in a similar social position. Edges with balance ratios far less than one may contain much more information of the network, which means a highly ranked user follows an ordinary user, reflecting some hidden real-world relationship or not apparent information between individuals. Therefore, research on edge balance ratio is of great significant on the microblogging platform.

As an indicator related only to local information, edge balance ratio can be easily calculated in distributed systems such as adaptive networks \cite{diffusion}, where lots of works have been done on the activities such as adaptive learning and  the diffusion process \cite{diffusionadaption,linkprobability,fast,adaptivelearning}. In these systems, edge balance ratio can be an important factor for both edges and vertices.

\subsection{Balance Profile and Positivity}

Since in-degree approximately reflects the importance of a vertex in the network, an edge with balance ratio larger than one can be called a \emph{positive edge}. An edge pointing from a high in-degree vertex to a low in-degree vertex is a \emph{negative edge}, correspondingly. An edge with balance ratio close to one can be called a \emph{normal edge}.

For the whole network, \emph{balance profile} is proposed as a global measure of the network, which is defined as the distribution of logarithmic edge balance ratio. The balance profile reveals the overall trend of the edges in the network. It reflects much more information, including the proportion of unbalanced edges of various levels.

On the basis of balance profile, the \emph{positivity} of a directed network can be defined as the expectation of logarithmic balance ratio with finite values. In a directed network with $N$ vertices, the logarithmic balance ratio of an edge ranges within $[-\log{(N-1)}, \log{(N-1)}]$. The expectation can be normalized by $\log{(N-1)}$,
\begin{align}
p =& \frac{1}{\log{(N-1)}}{\rm E}\{\log{R}\}\nonumber\\
 =& \frac{1}{|\mathcal E'|\log{(N-1)}}\sum_{({\rm A, B})\in {\mathcal E'}}\log{\frac{d_i({\rm B})}{d_i({\rm A})}},\nonumber
\end{align}
where $\mathcal E'$ is the set of edges with finite balance ratios.
The range of positivity is $[-1, 1]$.
As the average of logarithmic balance ratios for edges, the positivity reveals the positive level of the whole network. Especially, for a directed network with all the edges bi-directed, it can be calculated that the positivity is zero.

Positivity and assortativity reflect the properties of a directed network at different perspectives, though it seems that there are some similarities between positivity and in-in assortativity. Assortativity is a measure of the level that the vertices link to those with nearly the same degrees. However, positivity measures the level of vertices pointing to ones with higher degrees. An edge from a low in-degree vertex to a high in-degree vertex and an edge with the opposite direction both contribute a smaller assortativity to the network, while the former leads to a larger positivity and the latter leads to a smaller positivity. If the network has a large assortativity, the balance profile concentrates around the unitary edge balance ratio. Contrarily, the balance profile of a disassortative network is mainly far from the unitary balance ratio. The balance profile of a network with large positivity is mostly on the right side of the axis $R$ equals to one, while that of a less positive network has more on the left side.
Using both positivity and assortativity, the balance property of a network can be better described.

\begin{figure*}
\begin{center}
\centering
\includegraphics[width=\textwidth]{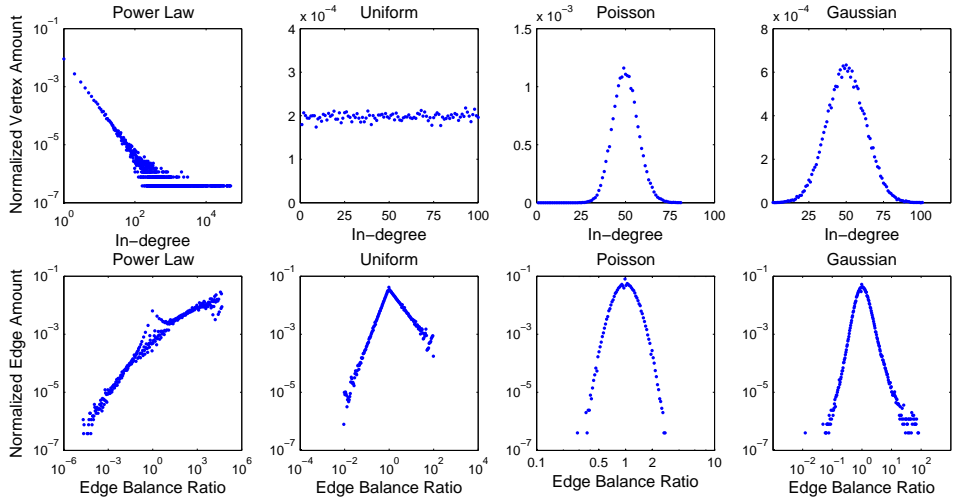}
\caption{The in-degree distributions and balance profiles of the four directed networks. The balance profiles have different shapes.}\label{fig1}
\end{center}
\end{figure*}

As an example, the in-degree distributions and balance profiles of four directed networks with different typical in-degree distributions are illustrated in Fig. \ref{fig1}. The four networks have different shapes of balance profiles. The positivities of the networks are $0.6474$, $0.0437$, $0.0019$ and $0.0068$, respectively. The power law network has a large positivity, because the positive half of the balance profile is significantly higher than the negative half. The balance profiles of the last two networks are almost symmetrical, which leads to rather small positivities.

\subsection{Basic Assumptions for the Network}
Power law distribution is the most common and the most important distribution in real-world networks.
Directed networks following power law are mainly discussed in this work.
According to the properties of real-world directed networks, two basic assumptions are adopted in the network model.

\begin{assumption}
The in-degrees of the vertices in the network follow power law distribution approximately,
$$
N_k\approx A k^{-\gamma},
$$
where $N_k$ is the amount of vertices with in-degree $k$, $A$ is a scale factor and $\gamma$ is the scaling exponent of power law.
\end{assumption}

\begin{assumption}
For any vertex ${\rm V}_0$,
$$
P(d_i({\rm V})=k | {\rm V}\in{\mathcal F}({\rm V}_0))\approx P(d_i({\rm V})=k),
$$
where $P(d_i({\rm V})=k)$ is the probability that vertex ${\rm V}$ has in-degree $k$ and ${\mathcal F}({\rm V}_0)$ is the set of followers of vertex ${\rm V}_0$.
\end{assumption}

Assumption 2 means that there is no bias on the followers of vertices on average. The proportion of various types of followers of a vertex approximately equals that of the whole network. The assumptions are reasonable in a microblogging network. Real data of microblogging network verifies the correctness of the assumptions.

The assumptions above are regarded as basic properties of the network. They are considered to be satisfied in the following of this paper.
\subsection{Theoretical Analysis on Edge Balance Ratio}
Firstly, the amount of edges with balance ratio $R$ and in-vertex degree $k$ is calculated, as Lemma 1. Please refer to the appendix for the proof of Lemma 1.

\begin{lemma}
If a power law directed network of $N$ vertices satisfies Assumptions 1 and 2, then the total amount of edges with out-vertices of in-degree $k$ and balance ratio $R$ is $\displaystyle{\frac{{A}^2}{N}k^{1-2\gamma}R^{\gamma}}$.
\end{lemma}

The main result on edge balance ratio is proposed as Theorem 1, which provides the approximate calculation of the edge balance ratio distribution.
Here, we consider the edges with finite balance ratios only.
\begin{theorem}
If an $N$-vertex power law directed network satisfies Assumptions 1 and 2, for logarithmically divided counting intervals
$$
[\cdots, \alpha^{-(s+1)}, \alpha^{-s}, \cdots,\alpha^{-2}, \alpha^{-1}, 1, \alpha, \alpha^2, \cdots, \alpha^s, \alpha^{s+1}, \cdots]
$$
of edge balance ratio,
the distribution of edge balance ratio follows power law piecewise.
In detail, the amount of edges with balance ratio $R$ satisfies
$$
N(R)\approx\displaystyle{\frac{{A}^2}{N}}\cdot\left\{
    \begin{array}{ll}
    \displaystyle{\frac{1-\alpha^{1-\gamma}}{(\gamma-2)(2\gamma-3)}R^{\gamma-1}}, &R\ll1; \\
    \displaystyle{\frac{1}{2\gamma-2}R^{\gamma}}, &R\lesssim1;\\
    \displaystyle{\frac{1}{2\gamma-2}R^{1-\gamma}}, &R\gtrsim1; \\
    \displaystyle{\frac{1-\alpha^{2-\gamma}}{(\gamma-2)(2\gamma-3)}R^{2-\gamma}}, &R\gg1.
    \end{array} \right.
$$
The scaling exponents of the four sections are $\gamma-1$, $\gamma$, $1-\gamma$ and $2-\gamma$, respectively, where $\gamma$ is the scaling exponent of the in-degree distribution of the network.
\end{theorem}

The outline of the proof is presented here while the details are included in the appendix. The cases of $R\ge1$ and $R<1$ are calculated separately.

Considering the edge balance ratios satisfying $R\ge1$, the axis is logarithmically divided into counting intervals by $[1, \alpha, \alpha^2, \cdots, \alpha^s, \alpha^{s+1}, \cdots]$. For each interval $[\alpha^s, \alpha^{s+1}]$, the contribution of edge balance ratios $R$ within the interval is the sum for all the combinations $(k,m)$ falling into this interval, where $m$ is the in-degree of the out-vertex and $k$ is the in-degree of the in-vertex. It should be noted that each value of $R$ is discrete. Not all values of edge balance ratio can have an edge exactly corresponding to it. The calculations for intervals with large $s$ and those with small $s$ are different. We divide the calculation into two parts, corresponding to the cases of $R$ far larger than one and slightly larger than one, respectively. For $R$ slightly larger than one, the sum vibrates strongly for various intervals. We focus on the peak values only and it follows the power law.

Similarly, considering the edge balance ratio less than one, the counting interval is $[\cdots, \alpha^{-(s+1)}, \alpha^{-s}, \cdots,$ $\alpha^{-2}, \alpha^{-1}, 1]$. For each edge balance ratio interval, we sum up all the combinations $(k,m)$ falling into it. The calculation is also divided into two parts. We focus on the peak values only for $R$ slightly smaller than one.

\subsection{Remarks on the Theoretical Analysis}
\begin{remark}
The theoretical analysis reveals that the edge balance ratio obeys the power law piecewise if the distribution of in-degree obeys the power law distribution. The scaling exponent of the distribution of edge balance ratio in each section is strictly linearly determined by the scaling exponent of in-degree distribution.
In-degree is the property of vertices while edge balance ratio is the property of edges. The relationship between the statistical properties of vertices and edges is established.
\end{remark}

\begin{remark}
Positive edges are much more than negative ones in a power law network.
In other words, there are much more edges with balance ratios larger than one than edges with balance ratios smaller than one. The trend of the overall network is positive, which means that most of edges in the network are positive or normal edges.

In Twitter and Sina Weibo, for a famous user, most edges related to it are from ordinary users, which leads to a large amount of positive edges. By contrast, for a user with few followers, most of his followers have similar ranks with him and there are few famous users follow him. Consequently, most edges pointing to him are normal edges.
\end{remark}

\begin{remark}
The distribution under logarithmical intervals in Theorem 1 is the balance profile of the network, which implies the characteristics of the network. The shape of the balance profile is determined by the scaling exponent $\gamma$.

The balance profile for $R<1$ always rises in the double logarithmic coordinate system.
According to the theoretical analysis, the slope is $\gamma$ when $R$ is slightly smaller than one while the slope is $\gamma-1$ when $R$ is far smaller than one. It is known that the scaling exponent $\gamma$ of a real-world network is always larger than one in general.

The shape of balance profile for $R>1$ is determined by the sign of $\gamma-2$.
The slopes are $1-\gamma$ and $2-\gamma$ for $R\gtrsim1$ and $R\gg1$, respectively. The balance profile always has a negative slope if $R$ is not far from $1$. For the case $\gamma$ larger than $2$, the profile has a negative slope for $R\gg1$. On the contrary, this segment of the curve will rise if $\gamma$ is less than $2$. In this case, edges with larger balance ratios account for a greater proportion. The critical value of $\gamma$ is $2$. It affects the trend of the balance profile a lot.
\end{remark}

\section{Numerical Simulations}
In this section, numerical simulations are done to verify the theoretical results. Real-world networks have a lot of randomness. In order to generate networks similar to real-world networks, two pairs of stochastic conditions are put forward, involving randomness into the network model. According to the conditions, four types of networks are constituted to simulate real-world networks.
The theoretical analysis is verified by simulations, including various stochastic networks with different power law scaling exponents.
The balance profiles are shown to be wizard-hat shaped.

\subsection{Network Models: Stochastic Conditions and Stochastic Networks}
In a real-world network, the basic assumptions are satisfied statistically but not strictly. In order to get a better simulation of real-world networks, randomness is involved in the network model.
The two pairs of conditions are as follows, for different interpretations of power law and the in-degree, respectively.
The total amount of vertices in the network is denoted as $N$ and $A$ is a scale factor.

$1$. The amount of the vertices with in-degree $k$ is $A\cdot k^{-\gamma}$;

$1'$. For any vertex, the event that its in-degree is $k$ occurs in probability $\displaystyle{\frac{A}{N}\cdot k^{-\gamma}}$.

$2$. A vertex has an in-degree $k$ means there are $k$ vertices pointing to it;

$2'$. A vertex has an in-degree $k$ means that any other vertex points to it with probability $k/N$.

The first pair of conditions describes definitions of the power law in different senses. Under the assumption of condition 1, the amount of vertices with in-degree $k$ is strictly determined. However, condition $1'$ involves randomness and conforms more to real-world networks. The second pair of conditions presents randomness in the definition of in-degree. Similarly, condition $2'$ reflects greater uncertainty, for the sake of matching the real-world networks. The two pairs of conditions describe the network from different aspects and they are independent of each other. The conditions within each pair are equivalent in an average sense, while $1'$ and $2'$ involve randomness to simulate real-world networks.

Utilizing the combinations of two pairs of stochastic conditions, four types of networks are defined as Table \ref{table1}. By the definition of deterministic network, the in-degree of vertices follows the power law strictly. The stochastic networks of type I and II involve the randomness in the power law distribution and the in-degree of vertices, respectively. The stochastic networks of type III involve the randomness of both two aspects. The four types of networks can be generated to simulate real-world networks.

\renewcommand\arraystretch{1.5}
\begin{table}
\begin{center}
\caption{Four Types of Networks.} \label{table1}
\begin{tabular}{c|cc}
\hline
& Condition $1$&Condition $1'$\\
\hline
Condition $2$ & Deterministic network & Type I stochastic network  \\
Condition $2'$ & Type II stochastic network & Type III stochastic network\\
\hline
\end{tabular}
\end{center}
\end{table}

\subsection{In-degree Distribution of Stochastic Networks}

\begin{figure*}
\centering
\includegraphics[width=\textwidth]{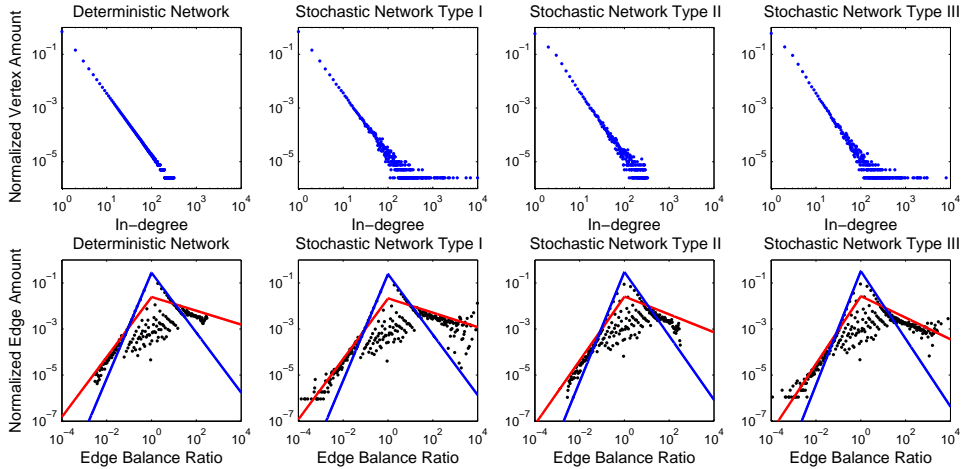}
\caption{In-degree and edge balance ratio distributions of four different types of stochastic networks with $\gamma=2.3$.}\label{fig2}
\end{figure*}

\begin{figure*}
\centering
\includegraphics[width=\textwidth]{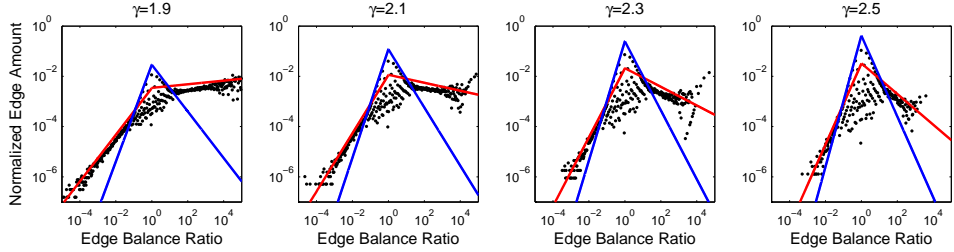}
\centering
\caption{The distributions of edge balance ratio for various scaling exponent $\gamma$ in stochastic network of type III.}\label{fig3}
\end{figure*}

Four networks with $400,000$ vertices are generated separately, where the scaling exponent of power law is $2.3$. The sub-figures at the top of Fig.~\ref{fig2} illustrates the in-degree distributions of four different types of networks defined previously.

It is indicated that randomness is involved in stochastic networks of type I, II, and III, while the in-degree distribution of the deterministic network is stepwise. Moreover, the largest in-degrees of deterministic network and stochastic network of type II are relatively small, which means there is no vertex being followed by large numbers of vertices in these two networks. However, vertices with large in-degrees appear frequently in real-world networks. In this perspective, stochastic networks of type I and III are more like real-world networks beyond the others.

\subsection{Numerical and Theoretical Results of Different Stochastic Networks}

The numerical and theoretical results are illustrated in the sub-figures at the bottom of Fig. \ref{fig2}. The networks are generated with $400,000$ vertices and scaling exponent $2.3$. Four sub-figures are for different stochastic networks described previously. The statistical results of balance profiles are shown as the points, while the four line segments indicate the theoretical results. Both theoretical and statistical results are wizard-hat shaped.

The balance profiles of four networks are almost the same with that of real-world networks.
For four different types of networks, the theoretical analysis matches the statistical result well. Four segments of power law compose the balance profile.

\subsection{Statistical and Theoretical Results of Different Scaling Exponents}

The scaling exponent $\gamma$ has a great influence on the property of networks. Power law networks with various scaling exponents $\gamma$ are generated to obtain various in-degree distributions. The stochastic network of type III is adopted in all of the simulations in this subsection. The statistical and theoretical results are illustrated in Fig. \ref{fig3}. The size of each network is $400,000$.

The statistical and theoretical balance profiles are indicated as the points and line segments in Fig. \ref{fig3}.
With a smaller $\gamma$, one flatter wizard-hat is got and with a larger $\gamma$, a sharper hat is obtained.
The statistical and theoretical results match well for various scaling exponent $\gamma$. Especially for the case where $\gamma$ is equal to $1.9$, the theoretical analysis shows that the slope of this segment should be $-0.1$. The statistical balance profile rises when the balance ratio is larger, actually the same with the theoretical result. This verifies the correctness of the theoretical analysis.

\section{Statistics of Real-world Networks}

\subsection{Datasets of Microblogging: Twitter and Sina Weibo}
Data of Twitter and Sina Weibo is used as the real-world network. In-degree distributions and edge balance distributions are obtained on the two datasets.

Both Twitter and Sina Weibo provide Application Programming Interfaces (API) for developers to collect data.
The dataset of Twitter is downloaded from \cite{whatis}. This dataset contains 35 million users and 1.4 billion
relationships. The dataset of Sina Weibo is rare and we use APIs to crawl what we need. The profile of a Sina
Weibo user includes user ID, screen name, gender, location and a brief description. Other data such as how
many statuses are published and who follows the user and whom the user follows are also listed. The crawl began in July, 2011 and by
now 110 million users and 6.96 billion relationships are gathered. The users we have crawled cover more than 30\% of all the users of Sina Weibo.

\subsection{In-degree Distributions and Balance Profiles of Twitter and Sina Weibo}

\begin{figure}
\centering
\includegraphics[width=4in]{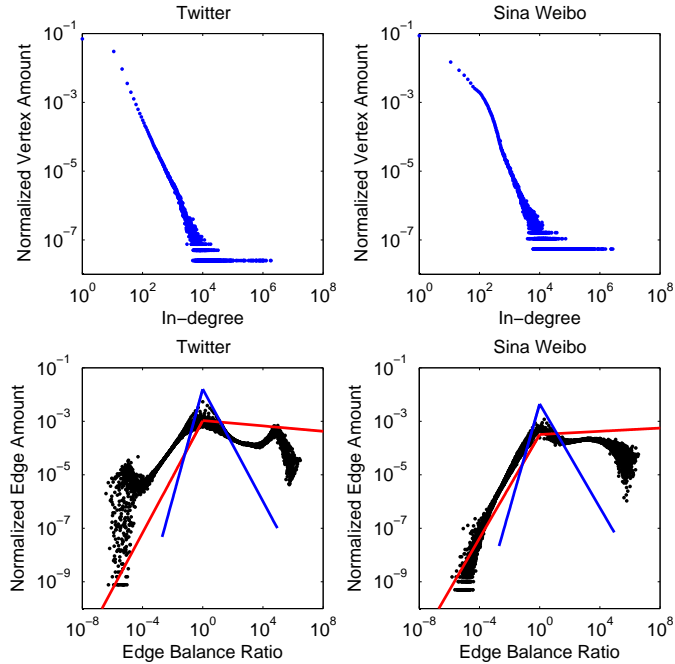}
\centering
\caption{The in-degree distributions and balance profiles of Twitter and Sina Weibo.}\label{fig4}
\end{figure}

The sub-figures at the top of Fig. \ref{fig4} illustrate the in-degree distributions of the datasets of Twitter and Sina Weibo. Both the distributions show the same characteristics of power law. The scaling exponent of Twitter is larger than that of Sina Weibo. The larger the scaling exponent is, the sparser the network is. The relationships between Sina Weibo users are much closer than that of Twitter. Due to Sina Weibo has a smaller scaling exponent, famous stars are more outstanding beyond ordinary users. The users of Sina Weibo are mostly from China while Twitter is used globally. That leads to that famous stars on Twitter are more dispersed than those on Sina Weibo. This matches the scaling exponents above.

The balance profiles of Twitter and Sina Weibo are illustrated in the sub-figures at the bottom of Fig. \ref{fig14}. Although the theoretical curves are similar with the  statistics of real-world networks, they are not strictly the same. The in-degree distributions of real-world networks do not follow the power law strictly, and the following relationships are not strictly sampled uniformly. These reasons lead to the errors. Especially for the edge balance ratio of Twitter, there is a peak for large balance ratio, which can hardly be predicted. The reason for the peak is that users with lower ranks are more interested in following the users with very high ranks, which is not strictly the same with Assumption 2 in the following relationship.

\section{Discussions}

\begin{figure*}
\begin{center}
\centering
\includegraphics[width=\textwidth]{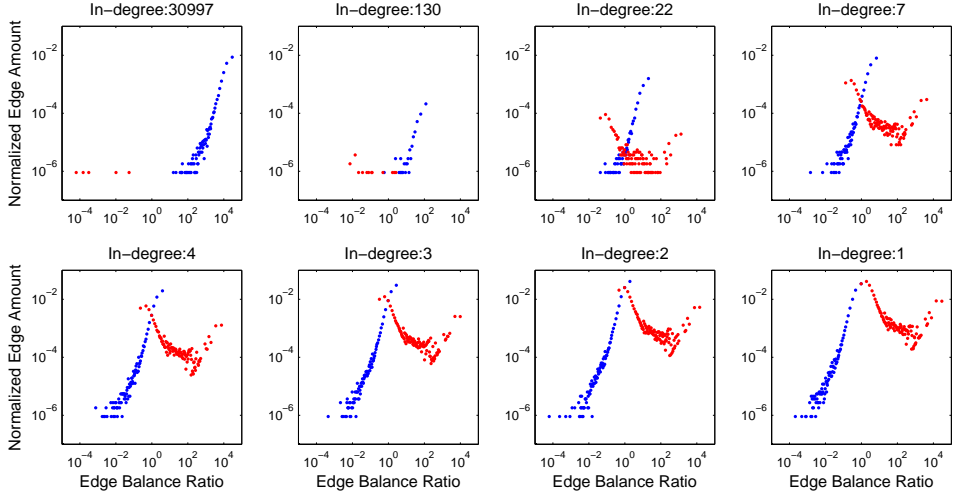}
\caption{The edge balance ratio distributions for edges related to vertices with in-degrees 30997, 130, 22, 7, 4, 3, 2, and 1.}
\label{fig5}
\end{center}
\end{figure*}
\subsection{Counting Intervals}
Related to the discrete property of edge balance ratio $R$ and the counting method, the cases of $R$ close to one and far from one are different.
For intervals near the unitary balance ratio, the counting intervals are short and only a few edge balance ratios fall into each of them. Therefore the edge balance ratios falling into different intervals are not even and the amount of edges vibrates a lot. For intervals far from the unitary balance ratio, large amounts of edges with various balance ratios fall into them. Thus statistical property affects a lot and the distribution looks smooth.

There is no strict threshold in the abscissa for cases of $R$ near one and far from one, which is determined by the interval parameter $\alpha$.
The larger the parameter $\alpha$ is, the larger the intervals are. More edges fall into each interval and the statistical property is more clearly shown. As a result, the section with the smooth statistical property is larger. On the contrary, the section with the vibrating property is larger when the interval parameter $\alpha$ is small.

\subsection{Error Analysis}
Although the theoretical slope in the double logarithmic coordinate system matches the simulation well, the intercept is not as precise, especially when $R$ is much larger than one.

The main error sources are in the following areas:

\begin{enumerate}
\item
Randomness is involved in the generated networks;

\item
The estimation for $A$ in the power law representation of $y$ equal $Ax^{-\gamma}$ is not precise enough;

\item
Some approximations are utilized in the theoretical analysis, such as replacing the summation with integration.
\end{enumerate}

These lead to the error of estimated intercept in the double logarithmic coordinate system.
However, the intercept is not as critical as the slope for the power law distribution. The scaling exponent is much more critical, which determines the property of the network.

\subsection{Further Investigations}

A stochastic network is generated with $400,000$ vertices and $\gamma$ equals $2.3$. The vertex with the largest in-degree has $30997$ followers.
If vertices are sorted by their in-degrees in descending order, the vertices at top $0.1\%$, $1\%$ and $5\%$ have $130$, $22$, and $7$ followers, respectively. Considering these vertices and vertices with in-degrees $4$, $3$, $2$, and $1$, the edges related to them are selected for further investigation. The balance ratios of these edges are shown in Fig. \ref{fig5}. Each sub-figure shows edges of all the vertices with the specified in-degree. The red points represent the edges start from the vertices, while the blue ones represent edges pointed to the vertices with the specified in-degrees.

The distributions of in-edges are similar for vertices with various in-degrees and so are the distributions of out-edges. The reason for this is the Assumption 2, which is that the in-degree distribution of the followers of a vertex is approximately the same as the distribution of all vertices in the whole network.

As the increasing of the in-degrees of vertices, the red points are moving towards left and the blue ones are moving towards right. This is because that for the red out-edges, the edge balance ratios are smaller if the in-degree of the vertex is larger. Similarly, for the blue in-edges, the balance ratios are larger if the vertex has a larger in-degree.
The wizard-hat-shape balance profile is the superposition of all the statistics like this.
The contribution of vertices with in-degrees $1$, $2$, $3$ and $4$ is critical, for the amount of these vertices is huge, although the followers of them are few. Correspondingly, the vertex with the most followers contributes a lot to the statistics of the balance ratio, even though there is only one such vertex in the network.

\begin{figure}
\centering
\includegraphics[width=4in]{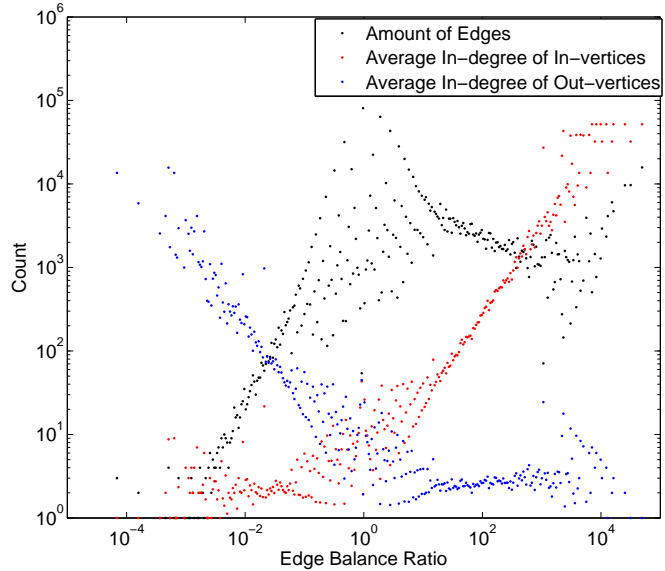}
\caption{Edge balance ratio distribution and average in-degrees of in-vertices and out-vertices.}\label{fig6}
\end{figure}

Fig. \ref{fig6} illustrates the average in-degrees of in-vertices and out-vertices of the edges in each counting interval. For edges with large balance ratios, the in-degrees of the out-vertices have no significant disparity, while the disparity between the in-vertices of these edges leads to the disparity of edge balance ratio. Correspondingly, for edges with small balance ratios, the disparity between the out-vertices contributes a lot. Therefore, an edge with a large balance ratio is mainly because the in-vertex has a large in-degree, while the reason for a very small edge balance ratio is that the out-vertex has a large in-degree.

\section{Conclusion}
In directed networks, edge balance ratio is an important measure of the balance property of edges, while balance profile is a description of the balance level of the network.
If the in-degree distribution follows the power law, the distribution of edge balance ratio follows a piecewise power law, which is wizard-hat shaped.
Numerical simulation results confirm the theoretical analysis. Real-world network datasets of Twitter and Sina Weibo are used to obtain the statistics of edge balance ratio, which is basically consistent with the theoretical results. Moreover, some related topics and detailed discussions are also included in this paper.

\appendix

\appendixtitleon
\begin{appendices}

\section{The Proof of Lemma 1}
By Assumption 1, there are $A\cdot k^{-\gamma}$ vertices with in-degree $k$ in the network. According to Assumption 2, among the followers of a vertex with in-degree $k$, there are $\frac{\textstyle k}{\textstyle N}\cdot A\cdot m^{-\gamma}$ vertices having $m$ followers. Consequently, the amount of edges with balance ratio $R$ whose target vertices having in-degree $k$ is
$$
\frac{k}{N}\cdot A\left(\frac{k}{R}\right)^{-\gamma}=\frac{A}{N}k^{1-\gamma}R^{\gamma},
$$
where
$$
R=\frac{k}{m}
$$
for the combination $(k,m)$.

Therefore, the number of vertices with in-degree $k$ and edge balance ratio $R$ is
$$
\frac{A}{N}k^{1-\gamma}R^{\gamma}\cdot Ak^{-\gamma}=\frac{{A}^2}{N}k^{1-2\gamma}R^{\gamma}
$$
altogether.

\section{The Proof of Theorem 1}
The results of four sections are proved separately, corresponding to different cases.
\subsection{The Case $R$ far larger than $1$}

For interval $[\alpha^s, \alpha^{s+1}]$ with a large $s$, the following inequality
$$
\alpha^s\le \frac{k}{m}\le\alpha^{s+1}
$$
is established for a fixed $k$.
Therefore, it satisfies
$$
\frac{k}{\alpha^{s+1}}\le m\le\frac{k}{\alpha^s}.
$$
Each pair of $(m,k)$ has a contribution of
$$
\frac{{A}^2}{N}k^{1-2\gamma}\left(\frac{k}{m}\right)^{\gamma}=\frac{{A}^2}{N}k^{1-\gamma}m^{-\gamma}
$$
edges within this interval.

For $k$ no less than $\alpha^{s+1}$, the amount of edges is approximately
$$
\frac{{A}^2}{N}k^{1-\gamma}\sum^{\lfloor\frac{k}{\alpha^s}\rfloor}
_{m=\lceil\frac{k}{\alpha^{s+1}}\rceil}m^{-\gamma}\approx\frac{{A}^2}{N}\frac{\alpha^{\gamma-1}-1}{\gamma-1}\alpha^{s(\gamma-1)}k^{2-2\gamma},\nonumber
$$
where the sum is approximated by the integration.
Summing over $k$, the result is approximately
\begin{equation}\label{eq1}
\;\frac{{A}^2}{N}\frac{\alpha^{\gamma-1}-1}{\gamma-1}\alpha^{s(\gamma-1)}
\sum_{k=\lceil\alpha^{s+1}\rceil}^{\infty}k^{2-2\gamma}
\approx\;\frac{{A}^2}{N}\frac{\alpha^{3-2\gamma}(\alpha^{\gamma-1}-1)}{(\gamma-1)(2\gamma-3)}\alpha^{s(2-\gamma)}.
\end{equation}

For $k$ satisfying $\alpha^s\le k\le\alpha^{s+1}$, the amount of edges in this interval is approximately
$$
\frac{{A}^2}{N}k^{1-\gamma}\sum^{\lfloor\frac{k}{\alpha^s}\rfloor}_{m=1}m^{-\gamma}
\approx\frac{{A}^2}{N}\frac{1}{\gamma-1}\left(k^{1-\gamma}-\alpha^{s(\gamma-1)}k^{2-2\gamma}\right).\nonumber
$$
Sum over $k$ and it leads to
\begin{equation}\label{eq2}
\frac{{A}^2}{N}\frac{1}{\gamma-1}\left(\sum_{\lceil\alpha^s\rceil}^{\lfloor\alpha^{s+1}\rfloor}k^{1-\gamma}
-\alpha^{s(\gamma-1)}\sum_{\lceil\alpha^s\rceil}^{\lfloor\alpha^{s+1}\rfloor}k^{2-2\gamma}\right)
\approx\frac{{A}^2}{N}\frac{1}{\gamma-1}\left(\frac{1-\alpha^{2-\gamma}}{\gamma-2}
-\frac{1-\alpha^{3-2\gamma}}{2\gamma-3}\right)\alpha^{s(2-\gamma)}.
\end{equation}

The sum of (\ref{eq1}) and (\ref{eq2}) is
$$
\frac{{A}^2}{N}\frac{1-\alpha^{2-\gamma}}{(\gamma-2)(2\gamma-3)}\alpha^{s(2-\gamma)}.
$$
It can be considered as the value at $\alpha^s$. In other words, the edge amount $N(R)$ with respect to balance ratio $R$ can be written as
$$
N(R)\approx\frac{{A}^2}{N}\frac{1-\alpha^{2-\gamma}}{(\gamma-2)(2\gamma-3)}R^{2-\gamma},
$$
which obeys the power law with scaling exponent $2-\gamma$.

\subsection{The Case $R$ slightly larger than $1$}

If the interval $[\alpha^s, \alpha^{s+1}]$ has a small $s$, the sum vibrates strongly for various intervals. We discuss the peak values only. Peak values are placed at integer edge balance ratios. According to the expression $\displaystyle{\frac{A^2}{N}k^{1-2\gamma}R^{\gamma}}$ of each $k$, the value at edge balance ratio $R$ is
\begin{equation}\label{eq3}
\frac{{A}^2}{N}\sum_{i=1}^{\infty}R^{\gamma}(iR)^{1-2\gamma}\approx\frac{{A}^2}{N}\frac{1}{2\gamma-2}R^{1-\gamma}.
\end{equation}
It demonstrates that the peak values obey the power law with scaling exponent $1-\gamma$.

Specially, the value $R=1$ corresponds to the case where the edge balance ratio is $1$.
By (\ref{eq3}), the amount of edges in this case is about $\displaystyle{\frac{{A}^2}{N}\frac{1}{2\gamma-2}}$.

\subsection{The Case $R$ far smaller than $1$}

For the interval $[\alpha^{-(s+1)}, \alpha^{-s}]$ with a larger $s$,
$m$ should satisfy the inequality
$$
\alpha^{-(s+1)}\le \frac{k}{m}\le\alpha^{-s}
$$
for a fixed $k$.
Therefore,
$$
k\alpha^{s}\le m\le k\alpha^{s+1}.
$$
The contribution of the combinations $(m,k)$ within this interval is
$$
\frac{{A}^2}{N}k^{1-2\gamma}\left(\frac{k}{m}\right)^{\gamma}=\frac{{A}^2}{N}k^{1-\gamma}m^{-\gamma}
$$
edges.
The amount of edges for $k$ is approximately
$$
\frac{{A}^2}{N}k^{1-\gamma}\sum^{\lfloor k\alpha^{s+1}\rfloor}_{m=\lceil k\alpha^{s}\rceil}m^{-\gamma}
\approx\frac{{A}^2}{N}\frac{1-\alpha^{1-\gamma}}{\gamma-1}\alpha^{s(1-\gamma)}k^{2-2\gamma}.\nonumber
$$
Then sum over $k$ and the result is
$$
\frac{{A}^2}{N}\frac{1-\alpha^{1-\gamma}}{\gamma-1}\alpha^{s(1-\gamma)}\sum_{k=1}^{\infty}k^{2-2\gamma}
\approx\frac{{A}^2}{N}\frac{1-\alpha^{1-\gamma}}{(\gamma-1)(2\gamma-3)}\alpha^{s(1-\gamma)}.\nonumber
$$

The sum above is the amount of edges at balance ratio $\alpha^{-s}$. It can be written as
$$
N(R)\approx\frac{{A}^2}{N}\frac{1-\alpha^{1-\gamma}}{(\gamma-2)(2\gamma-3)}R^{\gamma-1}.
$$
In other words, the number of edges obeys power law with scaling exponent $\gamma-1$.

\subsection{The Case $R$ slightly smaller than $1$}
For the interval $[\alpha^{-(s+1)}, \alpha^{-s}]$ with a small $s$, the edge amount value vibrates and the peak values are taken into consideration. Similarly, the peak values are at $\displaystyle{\frac{1}{k}}$ for each integer $k$.
According to the expression $\displaystyle{\frac{A^2}{N}k^{1-2\gamma}R^{\gamma}}$, the value of $R=\displaystyle{\frac{1}{k}}$ is
$$
\frac{{A}^2}{N}R^{\gamma}\sum_{i=1}^{\infty}i^{1-2\gamma}\approx\frac{{A}^2}{N}\frac{1}{2\gamma-2}R^{\gamma}.
$$
The peak values obeys the power law and the scaling exponent is $\gamma$.

When $R$ equals $1$, the above result is the number of edges with balance ratio equaling $1$. It is the same with the result in previous section.
\end{appendices}

\end{document}